
%
%
\newcount\eqnumber

\def\numbereq{\global\advance\eqnumber by 1 \eqno(\the\eqnumber)}
%
%
\newcount\refnumber

\def\tie{\noexpand~}
\immediate\openout1=refs.tex
\immediate\write1{\noexpand\frenchspacing}
\immediate\write1{\parskip=0pt}
\def\ref#1#2{\global\advance\refnumber by 1%
[\the\refnumber]\xdef#1{\the\refnumber}%
\immediate\write1{\noexpand\item{[#1]}#2}}

%
\def\sqr#1#2{{\vcenter{\vbox{\hrule height.#2pt
       \hbox{\vrule width.#2pt height#1pt \kern#1pt
               \vrule width.#2pt}
               \hrule height.#2pt}}}}
\def\Box{{\mathchoice \sqr65\sqr65\sqr33\sqr23}\,}

\def\mpl#1 {{\it Mod. Phys. Lett.\/ \bf A#1 }}
\def\pl#1 {{\it Phys. Lett.\/ \bf #1B }}
\def\np#1 {{\it Nucl. Phys.\/ \bf B#1 }}
\def\pr#1 {{\it Phys. Rev.\/ \bf D#1 }}
\def\prl#1 {{\it Phys. Rev. Lett.\/ \bf #1 }}
\def\ij#1 {{\it Int. Jour. of Mod. Phys.\/ \bf #1 }}
\def\cmp#1 {{\it Comm. Math. Phys.\/ \bf #1 }}
\def\s{(\sigma)}
\def\sp{(\sigma')}
\def\ov{\overline}
\def\Tb{\ov T}
\def\d{\delta(\sigma-\sigma')}
\def\dpr{\delta'(\sigma-\sigma')}

\def\dx{\partial X}
\def\dbx{{\overline\partial}X}
\def\dsx{\partial^2X}
\def\dbsx{{\overline\partial}^2X}
\def\knl{K^{\nu\lambda}}
\def\hnl{h^{\nu\lambda}}
\def\bnl{b^{\nu\lambda}}
\def\half{\textstyle {1\over 2}}
\def\eqalinno{{\global\advance\eqnumber by 1}(\the\eqnumber)}
\def\name#1{\xdef#1{\the\eqnumber}}
\baselineskip=14pt
\hsize=6in
\hoffset=.24in
\vsize=8.5in
\voffset=.25in
\parindent=3pc
\hfuzz=0.3pt

\rightline{\vbox{\hbox{RU-91-16}}}
\medskip
\rightline{November 1991}
\vskip .5in
\centerline{\bf ISOLATED STATES AND THE CLASSICAL PHASE SPACE}
\bigskip
\centerline{\bf OF 2-D STRING THEORY}
\vskip .7in
\centerline{\bf Mark Evans and Ioannis Giannakis\footnote{$^{\dag}$}
{\tenrm supported in part
by the Department of Energy Contract Number DOE-ACO2-87ER40325, TASK-B}}
\medskip
\centerline{\it The Rockefeller University}
\centerline{\it 1230 York Avenue}
\centerline{\it New York, NY 10021-6399}
\vskip .6truein
\centerline{\bf Abstract}
\medskip
{\narrower
\baselineskip=12pt
\tenrm We investigate the classical phase space of 2-d string theory. We derive
the linearised covariant equations for the spacetime fields by considering the
most general deformation of the energy-momentum tensor which describes $c=1$
matter system coupled to 2-d gravity and by demanding that it respect conformal
invariance. We derive the gauge invariances of the theory, and so investigate
the classical phase space, defined as the space of all
solutions to the equations of motion modulo gauge transformations. We thus
clarify the origins of two classes of isolated states.
 \bigskip}

\medskip
\line{\bf 1. Introduction \hfil}
\nobreak\bigskip
One interesting property of string theory is that, in addition to
the usual physical fields, it possesses so-called isolated states. Unlike
the usual particle-like excitations, these states exist only at particular
momenta, and may
\ref{\akov}{A.\tie Polyakov, \mpl 6 (1991), 635.}
be relics of a conjectured topological phase
\ref{\kra}{E.\tie Witten, \cmp 117, 353 (1988); \cmp 121, 351 (1989); \np 311
(1988), 46.}
in which the full symmetry of the theory is unbroken (an argument is given in
reference
\ref\dcftss{M.\tie Evans, B.\tie Ovrut, \pr41 (1990), 3149.}
as to why topological {\it world-sheet\/} field theories correspond
to such a symmetric phase).

Isolated states were first identified in the $c=1$ matrix model
\ref\gkln{D. Gross, I.R. Klebanov, M.J. Newman, \np 350 (1990), 621; D.Gross,
I.R. Klebanov, \np 359 (1991), 3.},
 and were then sought and found in a continuum analysis of $c=1$ matter coupled
to two dimensional gravity [\akov].
A careful analysis of the BRST cohomology
\ref\ian{B. Lian, G. Zuckerman, \pl 254 (1991), 417; \cmp 135, 547 (1991).}
identified these isolated physical states, as well as relatives at non-zero
ghost number.
More recently,
Witten \ref\gdring{E.\tie Witten, IAS report IASSNS-HEP-91/51 (August, 1991)}
has argued that much of the physics of the isolated states
is captured by a ring of dimension-zero operators, while Klebanov
and Polyakov \ref\rkp{I.\tie Klebanov and A.\tie Polyakov, Princeton
University report PUPT-1281 (September, 1991)} and Gross and
Danielson \ref\grod{D.\tie Gross and U.\tie Danielson,
Princeton University Report PUPT-1258 (1991)}
have studied their interactions.

The continuum world-sheet version of this model consists of a single scalar
field coupled  to two-dimensional gravity. Naively, this should correspond to
a string embedded in a one-dimensional target space. However, as is by now well
known
\ref\podj{J. Polchinski, \np 324 (1989)
, 123; S.R. Das, A. Jevicki, \mpl 5 (1990), 1639; A. Sengupta, S. Wadia,
\ij 6 (1991), 1961; S.R. Das, S. Naik, S.R. Wadia, \mpl 4 (1989), 1033.}
\ref\gkl{D. Gross, I.R. Klebanov, \np 352 (1990), 671.},
 the conformal degree of freedom of the world-sheet metric does not decouple
and gives
rise to a second dimension for the target space-time.

In two dimensions there are no transverse spatial
directions, and so we would naturally
expect that the only degree of freedom of the theory would be a massless,
``tachyon". However, it turns out that the higher excitations of the theory
cannot quite be gauged away and the isolated states are the remnants.
At mass-level one, the usual continuum argument for this [\akov]
starts from the physical state conditions
$$
\eqalignno{
p^{\mu}(p_{\mu}+2Q_{\mu}){\epsilon^{\lambda\nu}}&=0
&\eqalinno\name\eqclit\cr
(p_{\mu}+2Q_{\mu}){\epsilon^{\mu\nu}}&=0&\eqalinno\name\eqclanw\cr}
$$
where $p_\mu$ is the momentum of the state, $Q_\mu$ is the
``background charge,'' proportional to the gradient of the
dilaton, and $\epsilon^{\mu\nu}$ is the polarisation
of the state.
For generic momenta $p_\mu$, equations (\eqclit) and (\eqclanw)
are conditions on the polarisation $\epsilon^{\mu\nu}$,
but for the particular momentum
$p_{\mu}+2Q_{\mu}=0$, both conditions are satisfied for
arbitrary polarisation.
It is in this way that isolated states are usually identified.

While this argument is undoubtedly correct (it yields new BRST cohomology
classes [\ian] and singularities in matrix model amplitudes [\gkln]) it also
has
some puzzling features. Equation (\eqclit) is a mass-shell condition, but
equation (\eqclanw) is a space-time gauge condition. Selecting the gauge should
be a free choice of the physicist and devoid of physical content, yet it is the
degeneration of this gauge condition which relaxes the requirement of
transversality, and so yields new physical states.
For example, imposing instead the Hilbert gauge condition
$p_\mu\epsilon^{\mu\nu}+{1\over2}p^\nu\epsilon^\mu_\mu=0$
would not indicate anything of interest at $p_\mu=-2Q_\mu$.
Furthermore, the physical state conditions (\eqclit) and (\eqclanw)
are {\it not\/} sufficient by themselves to eliminate all degrees
of freedom, even at generic momenta: equation (\eqclit) is a
mass-shell condition, while equation (\eqclanw) enforces
transversality in the {\it two\/}-dimensional
space-time. Obviously, $\epsilon=0$ is not the only solution
to equation (\eqclanw). To eliminate these degrees of freedom
it is necessary to invoke additional gauge invariances not manifest
in either (\eqclit) or (\eqclanw). The origin of the states is
thus rather unclear from (\eqclit) or (\eqclanw) alone.

Isolated states appear in another context in
\ref\gcovdef{M.\tie Evans, I.\tie Giannakis, \pr44 (1991), 2467.}.
 The present authors took critical string theory, and considered the most
general infinitesimal deformation of naive dimension two that preserves
conformal invariance. It was found that there is a distinct such
deformation for every solution to a set of linearized,
gauge covariant space-time
equations of motion for the massless states of the theory. However,
over and above this there remains a finite-dimensional space of
{\it additional\/} deformations,
which, it was argued, correspond to adding to the world-sheet action
terms proportional to instanton
topological charges. If these deformations are
integrable (can be made finite rather than just infinitesimal---a non trivial
question) they can be thought of as free parameters of string
{\it theory\/} itself
(rather than just moduli of the solution space). However, they may equally well
be thought of as corresponding to isolated states, a fact which may be seen as
follows. To each conformal field theory there corresponds a nilpotent BRST
operator $Q_B$, and a conformal deformation therefore corresponds to a
deformation of $Q_B$ which preserves nilpotency,
$Q_{B}^{2}={(Q_B+{\delta}Q_B)}^
{2}=0$. Thus $\{{Q_B,{\delta}Q_B}\}=0$, which implies that
${\delta}Q_B$ acting on one physical state yields another. Note that
this argument holds irrespective of the integrability of the conformal
deformation.

These isolated states (in the critical case, at zero momentum, as they must be
since Lorentz invariance is unbroken) are thus directly associated with global
aspects of the embedding of the string in space-time, once again suggesting a
connection with a topological phase of the theory.

There are certain similarities between the two types of isolated state; both
involve additions to the world-sheet energy-momentum tensor which are total
derivatives, and so, ``topological" in nature. It is therefore desirable to
have
a formalism in which both types of isolated states may be examined, and which
clarifies the origin of the first class without appealing
to the degeneration of a gauge condition.
We shall exhibit such a formalism by using the method of
[\gcovdef] applied to the case of two-dimensional
non-critical string theory.
By considering infinitesimal conformal deformations of the
world-sheet energy-momentum tensor we shall obtain the fully
gauge covariant linearised equations of motion for the space-time
fields, and by discussing the appropriate inner automorphisms
we shall derive the gauge invariances of the {\it interacting\/}
theory.

Together, these two pieces of information are sufficient for us to
construct (locally) the classical phase space of
the theory in an elementary
way; we simply fix the gauge of the solutions until there is
no gauge freedom left. The classical phase space is then isomorphic
to this gauge-fixed solution-space. In this way, both classes
of isolated state can be seen straightforwardly, and we do not
have to appeal to the degeneration of one particular gauge
condition to see them.

\bigbreak\medskip
\line{\bf 2. Conformal Deformations and Equations of Motion \hfil}
\nobreak\bigskip

In the continuum language, the two-dimensional string is usually (and
most simply) described by the world-sheet energy-momentum tensor
$$
\eqalign
{T({\sigma})&=\textstyle{1\over 2}{\eta^{\mu\nu}}
{\partial}{X_{\mu}}{\partial}{X_{\nu}}
{({\sigma})}+{Q_{\mu}}{\partial^{2}}{X^{\mu}}{({\sigma})}\cr
\overline T({\sigma})&=\textstyle{1\over 2}{\eta^{\mu\nu}}
{\overline{\partial}}{X_{\mu}}{\overline{\partial}}{X_{\nu}}
{({\sigma})}+{Q_{\mu}}{\overline{\partial}^{2}}{X^{\mu}}{({\sigma})}\cr}
\numbereq\name{\eqcurved}
$$
where $Q_{\mu}={(Q,0)}$ and $X^\mu$ stands for the two space-time
coordinates $(\phi, X)$.

Since moments of this choice of $T$ and $\Tb$ give two mutually
commuting copies of the Virasoro algebra, they define a conformal
field theory \ref\sov{A. Belavin, A.M. Polyakov, A. Zamolodchikov,
\np 241 (1984), 333; D.\tie Friedan, E.\tie Martinec, S.\tie Shenker,
\np 271 (1986), 93.}, and the central charge is $c=2+12Q^2$.
Thus a sensible string theory requires that $Q=\sqrt 2$
\ref\dkd{J. Distler and H. Kawai, \np 321 (1989), 509; F.\tie David, \mpl 3
(1988), 1651.}.

Conformal field theories (with the correct central charge) are solutions to
the classical equations of motion of string theory \ref\crap{C.\tie Lovelace,
 \pl 135 (1984), 75; C.\tie Callan, D.\tie Friedan, E.\tie Martinec, M.\tie
Perry, \np 262 (1985), 593; A.\tie Sen, \pr 32 (1985), 2102.}.
Thus we may obtain the linearised space-time equations of motion
by considering infinitesimal deformations of $T$ and $\Tb$ that
preserve the two mutually commuting Virasoro algebras [\dcftss],
\ref\imls{M. Evans and B. Ovrut, \pl231 (1989), 80.}
(see also \ref\byn{T. Banks, D. Nemeschansky and A. Sen, \np 277 (1986), 67.}).
The variations in $T$ and $\Tb$ must therefore
satisfy the deformation equations
$$
\eqalignno{[\delta T\s, T\sp]+[T\s, \delta T\sp]&=2\imath\delta T\sp\dpr-
\imath\delta T^{\prime} \sp\d&{\global\advance\eqnumber by 1} (\the\eqnumber a)
\name{\eqdif} \cr
[\delta\Tb\s, \Tb\sp]+[\Tb\s, \delta \Tb\sp]&=-2\imath\delta \Tb\sp\dpr+
\imath\delta \Tb^{\prime} \sp\d&(\the\eqnumber b) \cr
[\delta T\s, \Tb\sp]+[T\s, \delta\Tb\sp]&=0&(\the\eqnumber c) \cr}
$$

The usual physical state conditions (equations (\eqclit) and (\eqclanw))
may be obtained by considering {\it canonical\/} deformations
[\dcftss], [\imls], \ref\phil{M.\tie Campbell, P.\tie Nelson,
E.\tie Wong, University of Pennsylvania Report UPR-0439T (1990)},
which are defined by $\delta T\s=\delta\Tb\s=\Phi_{(1,1)}\s$, a primary
field of dimension (1,1). For a deformation corresponding to a state
at the first mass-level, such a deformation would be of the form
$$
{\delta}{T\s}={\delta}{\Tb\s}={h^{\mu\nu}}{(X)}{\partial}{X_{\mu}}{\ov\partial}
{X_{\nu}}\numbereq\name{\eqham}
$$
where the right hand side of this equation is a $(1,1)$ primary field only
if $h^{\rho\sigma}$ obeys
$$
\eqalign
{{({\Box}+2{Q_{\rho}}{\partial^{\rho}})}{h^{\mu\nu}{(X)}}&=0\cr
{({\partial_{\mu}}+2{Q_{\mu}})}{h^{\mu\nu}{(X)}}&=0\cr}\numbereq\name{\eqglo}
$$
which are obviously the same as equations (\eqclit) and (\eqclanw).

However, as was shown in [\gcovdef], canonical deformations are
not the most general. We may find deformations that replace the equation
of motion and gauge condition of equations (\eqglo) by a gauge-covariant
equation of motion alone. The derivation is very similar to that in
[\gcovdef], so we shall only outline it here, and refer the interested
reader to [\gcovdef] for more details.

We shall consider the most general local deformation which only contains terms
of naive dimension two for the graviton, dilaton and
antisymmetric tensor fields and
a term of naive dimension zero for the tachyon field, and write it as
$$
\eqalign{\delta
T&=H^{\nu\lambda}(X)\dx_\nu\dbx_\lambda+A^{\nu\lambda}(X)\dx_\nu
\dx_\lambda\cr
&\hphantom{=H^{\nu\lambda}(X)\dx_\nu\dbx_\lambda}\quad
+B^{\nu\lambda}(X)\dbx_\nu\dbx_\lambda
+C^\nu(X)\dbx_\nu+D^\lambda(X)\dbsx_\lambda+{\cal T}(X)\cr
\noalign{\vskip 2\jot}
\delta \Tb&=\ov H^{\nu\lambda}(X)\dbx_\nu\dx_\lambda+\ov A^{\nu\lambda}(X)
\dbx_\nu
\dbx_\lambda\cr
&\hphantom{=\ov H^{\nu\lambda}(X)\dbx_\nu\dx_\lambda}\quad
+\ov B^{\nu\lambda}(X)\dx_\nu\dx_\lambda
+\ov C^\nu(X)\dbsx_\nu+\ov D^\lambda(X)\dsx_\lambda+
{\cal\ov T}(X)\cr}\numbereq\name\eqansatz
$$
The tensors $H^{\nu\lambda}\ldots{\cal\ov T}$ are initially taken to be
completely independent. This {\it ans\"atz\/} is then substituted into the
deformation equations (\eqdif) yielding fifty-four
distinct equations for the ten
tensors $H^{\nu\lambda}\ldots\ov D^{\lambda}$. After some calculation these
equations reduce to the following
$$
\eqalign{\delta T\s &= \knl \dx_\nu \dbx_\lambda+ \left(\partial-\ov\partial
\right) [C^\nu\dx_\nu-D^\lambda\dbx_\lambda]+{\cal T}(X)\cr
\delta \Tb\s &= \knl \dx_\nu \dbx_\lambda-\left(\partial-\ov\partial\right)
[\ov C^\nu\dbx_\nu-\ov D^\lambda\dx_\lambda]+{\cal T}(X)\cr}
\numbereq\name\eqsoln
$$
where we have introduced a new field $K^{\nu\lambda}$, defined by
$$
K^{\nu\lambda}=H^{\nu\lambda}+{\partial^{\lambda}}{C^{\nu}}+{\partial^{\nu}}
{D^{\lambda}}\numbereq\name{\eqmucha}
$$
The quantities $C^{\mu}, {\ov C}^{\mu}, D^{\mu}, {\ov D}^{\mu}$ are given in
terms of $K^{\mu\nu}$ by\goodbreak
$$
\eqalignno{\partial_\nu C^\nu+2Q_\nu C^\nu&=0&{\global\advance\eqnumber by 1}
(\the\eqnumber a)\name{\eqd}\cr
D^\lambda&=-\half\partial_\mu K^{\mu\lambda}-Q_\mu K^{\mu\lambda}
&(\the\eqnumber b)\cr
\ov D^\lambda&=-\half\partial_\mu K^{\lambda\mu}-Q_\mu K^{\lambda\mu}
&(\the\eqnumber c)\cr
\partial^\lambda C^\nu&=\half\Box\knl+Q_\mu{\partial^{\mu}}\knl-\half\partial^
\nu\partial_\mu K^{\mu\lambda}-Q_\mu{\partial^{\nu}} K^{\mu\lambda}
&(\the\eqnumber d)\cr
\partial^\lambda\ov C^\nu&=\half\Box K^{\lambda\nu}
+Q_\mu{\partial^{\mu}} K^{\lambda\nu}-\half\partial^
\nu\partial_\mu K^{\lambda\mu}-Q_\mu{\partial^{\nu}} K^{\lambda\mu}
&(\the\eqnumber e)\cr}
$$
while the equation of motion for the tachyon field is given by
$$
\Box {\cal T} +2Q_{\mu}{\partial^{\mu}}{\cal T}+2{\cal T}=0\numbereq
\name\eqtach
$$
At first sight there is no equation of motion for the physical field
$K^{\mu\nu}$. However, just as in the critical case [\gcovdef],
the last two equations cannot be solved for $C^{\mu}$ and
$\ov C^{\mu}$ for arbitrary $K^{\mu\nu}$. There is an integrability condition
which $K^{\mu\nu}$ must satisfy which turns out to yield an equation of motion
$$
\Box\hnl+2Q_\mu{\partial^{\mu}}\hnl-\partial^\nu\partial_\mu h^{\mu\lambda}
-2Q_{\mu}{\partial^{\nu}} h^{\mu\lambda}-\partial^\lambda\partial_\mu h^{\mu\nu
}-2Q_{\mu}{\partial^{\lambda}} h^{\mu\nu}+\partial^{\nu}\partial^{\lambda} h
=\partial^{\nu}\partial^{\lambda}{\phi}\numbereq\name{\eqteta}
$$
for some scalar function $\phi(X)$ which we identify as the dilaton.
Here $h^{\mu\nu}$, which we identify as the graviton, is the symmetric
part of $K^{\mu\nu}$.

For the two-form field $b_{\mu\nu}$, the anti-symmetric part of
$K_{\mu\nu}$, the fact that the space-time is two-dimensional makes
for some significant differences from the critical case. As in the
critical case, the general solution to the integrability condition
involves a constant two-form, $\alpha_{\mu\nu}$, which makes its
appearance in the equation of motion for $b_{\mu\nu}$:
$$
\partial^\mu H_{\mu\nu\lambda} + 2Q^\mu H_{\mu\nu\lambda}
= \alpha_{\nu\lambda}
$$
where $H_{\mu\nu\lambda}=\partial_{[\mu}b_{\nu\lambda]}$ and
square brackets imply anti-symmetrisation. However, in two dimensions
$H$ vanishes identically, so that we have to take $\alpha=0$, and
$b_{\nu\lambda}$ is not constrained by any equation of motion.

Superficially, this might suggest that the isolated states of the type
found in [\gcovdef] do not occur in two dimensions (in the critical
case they were associated with the constants $\alpha_{\nu\lambda}$).
However, in section 4 we shall see that the difference is small, and
that the same isolated states occur in both theories.

It only remains to derive the equation of motion for the dilaton. The
integrability condition for $K^{\mu\nu}$ provides an expression for the
$C^{\nu}$ field, which is
$$
2C^{\nu}={\partial^{\nu}}(\phi-K)+{\partial_{\mu}}{K^{\nu\mu}}
+2Q_{\mu}K^{\nu\mu}\numbereq\name\eqstrongo
$$
By substituting this expression into the equation (\eqd a) and using the
equation of motion for the $h^{\mu\nu}$ field we derive the following equation
of motion for the dilaton field,
$$
\Box\phi+4Q_{\mu}{\partial^{\mu}}{\phi}=2Q_{\mu}{\partial^{\mu}}h
-4Q_{\mu}{\partial_{\nu}}h^{\mu\nu}-8Q_{\mu}Q_{\nu}h^{\mu\nu}\numbereq\name\eqlo
$$
We have thus achieved our goal of finding deformations which are associated
with covariant equations of motion and no gauge conditions.
This will enable us to investigate the isolated states
after we have understood the gauge symmetries of the theory.
It is to this question that we now turn.

\bigbreak\medskip
\line{\bf 3. Symmetries \hfil}
\nobreak\bigskip
In this section we shall use the method developed in [\dcftss],
[\gcovdef], [\imls] to derive the
gauge invariances of string theory around flat two-dimensional
space-time in the presence of a dilaton background linear in $\phi$.
The basic idea is that a symmetry transforms one solution of the
classical equations of motion to another, without changing the
physics. In the case of string theory, we are therefore interested
in physically indistinguishable conformal field theories.

Given a conformal field theory, one simple way to get another which
is physically indistinguishable is to take the operators of the
given theory and perform the same similarity transformation on them
all. This gives a new conformal field theory which is physically
identical to the old one, since none of the algebraic properties
are changed by a similarity transformation. However, in general
the energy-momentum tensors are different fields, and if that
change can be interpreted as a change in the space-time fields,
then that change is a symmetry transformation.

If we restrict ourselves to infinitesimal transformations, then
$\Phi\rightarrow\Phi+\delta\Phi$ is a symmetry if there exists
an operator $h$ such that
$$
i[h, T_\Phi\s]=T_{\Phi+\delta\Phi}\s-T_\Phi\s \numbereq\name\eqh
$$

Which operators $h$ satisfy equation (\eqh),
when $T_\Phi$ is the energy-momentum
tensor that describes the $c=1$ matter system coupled to two-dimensional
gravity? Because the commutator preserves naive dimension [\gcovdef],
the generators of interest are
$$
h={\int d{\sigma}}(\xi^{\mu}(X){\partial}X_{\mu}+\zeta^{\mu}(X){\ov\partial}
X_{\mu})\numbereq\name{\eqkormos}
$$
Now since
$$
\eqalign{\delta T\s&=-\imath [h, T\s]\cr
\delta\Tb\s&=-\imath [h, \Tb\s]\cr}\numbereq\name{\eqmala}
$$
is an automorphism, it automatically satisfies the deformation equations
and so must correspond to a solution of the type specified in equations
(\eqsoln) and (\eqd).
 Let
us therefore compute the right hand side of equation (\eqmala),
with $h$ given by (\eqkormos).
The result provides us with the transformation properties of the physical
fields
under coordinate and two-form gauge transformations
$$
\eqalignno{
\knl&=H^{\nu\lambda}+{\partial^{\lambda}}{C^{\nu}}+{\partial^{\nu}}{D^{\lambda}}
={\partial^{\lambda}}{\xi^
{\nu}}+{\partial^{\nu}}{\zeta^{\lambda}}&{\global\advance\eqnumber by 1}
(\the\eqnumber a)\name{\eqputa}\cr
\partial^{\nu}{\phi}&=-2Q_{\mu}{\partial^{\nu}}{\xi^{\mu}}
-2Q_{\mu}{\partial^{\nu}}{\zeta^{\mu}}&(\the\eqnumber b)\cr}
$$
These are the canonical general-coordinate and two-form gauge
transformations when the background consists of a flat space-time
and a non-constant dilaton. Thus, having obtained the linearised
equations of motion and gauge symmetries, we now proceed to the
construction of the classical phase space.

\bigbreak\medskip
\line{\bf 4. Classical Phase Space \hfil}
\nobreak\bigskip
We proceed now to investigate the classical phase space of our theory. It is
defined as the space of all solutions of the classical equations of motion
modulo gauge transformations.

As we saw in the last section,
the equations of motion for the physical fields are invariant under
two-dimensional coordinate transformations
$$
\hnl
\rightarrow{\hnl+\half{\partial^{\nu}}u^{\lambda}+\half{\partial^{\lambda}}
u^{\nu}}\numbereq\name{\eqyia}
$$
where $u^\lambda=\xi^{\lambda}+\zeta^{\lambda}$, and two-form gauge
transformations
$$
\bnl
\rightarrow{\bnl+\half{\partial^{\nu}}w^{\lambda}-\half{\partial^{\lambda}}
w^{\nu}}\numbereq\name{\eqdim}
$$
where $w^\lambda=\xi^{\lambda}-\zeta^{\lambda}$, while the dilaton transforms
in a non-trivial manner
$$
\partial^{\nu}{\phi} \rightarrow {\partial^{\nu}{\phi}-2Q_{\mu}{\partial^{\nu}}
u^{\mu}}\numbereq\name{\eqstella}
$$

Let us begin with the graviton and dilaton.
We find it convenient to impose first the Lorentz gauge condition
$$
Q_{\mu} h^{\mu\nu}=0 \numbereq\name{\eqvivi}
$$
where $Q_{\mu}=(Q,0)$.
Thus  equation (\eqvivi) is equivalent to the two conditions
$h^{00}=0$ and $h^{01}=0$, leaving $h^{11}$ as the sole non-zero
component. It is straightforward to see that we may impose this
condition, starting from an arbitrary $h^{\mu\nu}$; we need to find
$u^\mu$ which satisfies
$$
\eqalignno{\partial_0u_0&=-h_{00}&\eqalinno\name\equo\cr
\partial_0u_1&=-2h_{01}-\partial_1u_0&\eqalinno\name\equl\cr}
$$
Equations (\equo) and (\equl), thought of as equations for $u_0$
and $u_1$ respectively, obviously have solutions for arbitrary
right-hand sides.

In this gauge the equations for the graviton-dilaton system reduce to the
following set of equations
$$
{\partial_{0}}{\partial^{0}}{(\phi-h)}=0, \qquad h=h^{11} \numbereq\name{\eqwe}
$$
$$
{\partial^{0}}{({\partial_{0}}+2Q_{0})}h={\partial_{1}}{\partial^{1}}{\phi}
\numbereq\name{\eqkula}
$$
$$
{\partial^{0}}{\partial_{1}}{\phi}=0 \numbereq\name{\eqakis}
$$
$$
{\partial^{0}}{\partial_{0}}{\phi}+{\partial^{1}}{\partial_{1}}{\phi}+
4Q_{0}{\partial^{0}}{\phi}=2Q_{0}{\partial^{0}}h \numbereq\name\eqprotasov
$$
Although we have imposed a gauge condition there is still residual coordinate
invariance,
 namely coordinate transformations which respect this particular choice. These
are generated by those vector fields $u^{\mu}=(u^{0},u^{1})$ which obey the
following relations as can be seen by equation (\eqyia)
$$
{\partial_{0}}{u^{0}}=0, \qquad {\partial_{1}}{u^{0}}={\partial_{0}}{u^{1}}
\numbereq\name{\eqspar}
$$
with general solution
$$
u^0=u^{0}(x^{1}), \qquad u^{1}({x^{0}},{x^{1}})=({\partial_{1}}{u^{0}})x^{0}
+f(x^{1}) \numbereq\name{\eqlakis}
$$
where $f=f(x^{1})$ is an arbitrary function of $x^1$.

To fix the gauge further, we must make use of the equations of motion.
The most general solution to the equation (\eqakis) is
$$
\partial_{1}{\phi}=y(x^{1}) \numbereq\name{\eqmakis}
$$
where $y=y(x^{1})$ is an arbitrary function of $x^1$. According to (\eqstella)
$\partial_{1}{\phi}$ transforms under a coordinate transformation as
$$
\partial_{1}{\phi} \rightarrow{\partial_{1}{\phi}}-2Q_{0}{\partial_{1}}u^{0}
$$
We choose $u^{0}(x^1)$ so that $2Q_{0}{\partial_{1}}u^{0}=y(x^{1})$
and so set
$\partial_{1}{\phi}$ equal to zero.
Thus the gauge freedom associated with $u^{0}$ has been exhausted.
At our disposal we
still have the transformations which are generated by $u^{1}=u^{1}(x^{1})$.

Using $\partial_1\phi=0$ and equations (\eqwe), (\eqkula) and
(\eqprotasov) it is straightforward to show that
$$
\phi(x^{0})-h(x^{0},x^{1})=\rho(x^{1}) \numbereq\name{\eqsakis}
$$
where  $\rho(x^{1})$ is an arbitrary function of $x^{1}$ only.
 Using coordinate transformations which are generated by $u^{1}=u^{1}(x^{1})$
we can set $\rho(x^{1})$ equal
to zero, and all gauge freedom is now used up.
We are thus reduced to a single degree of freedom depending on
$x^0$ alone: $\phi(x^0)=h(x^0)$.
Substituting this degree of freedom into the equations of motion,
we find the following general solution:
$$
h(x^0)=\phi(x^{0})=-{{c_1}\over 2Q_{0}}
{\exp {(-2Q_{0}x^{0}})}+c_2 \numbereq\name{\eqpopi}
$$
where $c_1$, $c_2$ are arbitrary integration constants.

These, then, are the first class of isolated states.
Recalling that these linearised solutions are the
wave functions for the graviton and dilaton vertex operators, these new factors
might have been anticipated since the effective string coupling constant is
${\exp({\phi})}={\exp({-2Q_{0}x^{0}})}$.

Higher massive fields are expected to contribute in an analogous
manner to the phase space of the theory, namely states which
correspond to
discrete values of momenta. Consequently our phase space consists of one field,
the tachyon, and a tower of states which are remnants of the higher
massive modes of the string.

Finally, we turn to the two-form field, $b_{\mu\nu}$. Recall that
this field is not constrained by any equation of motion, and is
invariant under the gauge transformation, equation (\eqdim).
Thus, in two dimensions the two-form is always locally pure gauge
(recall that there are no three-forms, so every two-form is closed):
we do not even have to use an equation of motion to eliminate all
propagating degrees of freedom. However, depending on the global
properties of space-time, there may be cohomology. For example,
if space-time is compact, $b_2=1$ and there is a one dimensional
space of gauge-inequivalent solutions.

Although the derivation is superficially different, we wish to argue
that these states are the two-dimensional analogues of the second
class of isolated states found in [\gcovdef]. The reason is that
non-trivial cohomology is associated with non-trivial homotopy
\ref\bott{R. Bott and L. Tu, {\it Differential Forms and Algebraic
Topology}, Springer Verlag (1983).},
which in turn implies the existence of instantons.
Indeed, the two-form $b$ is just proportional to the instanton
topological charge density.
It was argued in [\gcovdef] that the isolated states found there
were associated with instanton topological terms in the action,
of just this type, which correspond to two-forms representing
non-trivial elements of $H^{(2)}$. Thus the absence of the parameters
$\alpha_{\mu\nu}$ in two dimensions is not due to the absence of the
isolated states, but rather to the absence of a propagating two-form,
so that there is no equation of motion to be modified.

We have thus succeeded in exhibiting both classes of isolated states.

\bigbreak\medskip
\line{\bf 6. Conclusions \hfil}
\nobreak\bigskip

In summary, we have used the techniques of [\gcovdef] to derive
isolated states of string theory in a very physical way; we obtained
linearised equations of motion and the gauge symmetries, and then
derived the phase space as the set of all solutions, modulo gauge
transformations.

\bigbreak\medskip
\line{\bf Acknowledgements. \hfil}
\nobreak\bigskip

I.G. would like to thank A. Polychronakos and M.E. would like to
thank D. Gross and S. Wadia for stimulating discussions. M.E. would
also like to thank Prof. Y. Frishman and the Weizmann Institute for
hospitality.

\immediate\closeout1
\bigbreak
\line{\bf References. \hfil}
\nobreak\medskip\vskip\parskip

\input refs

\vfill\end